\documentclass[journal,twocolumn]{IEEEtran}
\usepackage{amsfonts}
\usepackage{times}
\usepackage{graphicx}
\usepackage{latexsym}
\usepackage{dsfont}
\usepackage{amssymb}
\usepackage{amsmath}
\usepackage{cite}
\usepackage{verbatim}
\usepackage{subfigure}

\def\bb0{{\mathbb{0}}}

\def\ba{{\mathbf{a}}}
\def\bb{{\mathbf{b}}}

\def\bff{{\mathbf{f}}}

\def\b0{{\mathbf{0}}}

\def\sf0{{\mathsf{0}}}

\usepackage{epstopdf}
\usepackage{enumerate}
\usepackage{algorithmicx}
\usepackage{algorithm}
\usepackage{amsmath}
\usepackage[noend]{algpseudocode}
\usepackage{float}
\usepackage{hyperref}
\usepackage{color}
\usepackage{makeidx}
\usepackage{bbm}
\usepackage{graphicx}
\usepackage{lipsum}
\usepackage{subfigure}
\usepackage{tablefootnote}
\usepackage{multirow}
\usepackage{multicol}
\usepackage{balance}

\def\j{\mathrm{j}}

\newcommand\Tstrut{\rule{0pt}{2.6ex}}

\begin{document}

\title{ViWi Vision-Aided mmWave Beam Tracking: Dataset, Task, and Baseline Solutions}
\author{Muhammad Alrabeiah, Jayden Booth, Andrew Hredzak, and Ahmed Alkhateeb \\ Arizona State University, Emails: \{malrabei, jayden.booth, ahredzak, alkhateeb\}@asu.edu}
\maketitle

\begin{abstract}
Vision-aided wireless communication is motivated by the recent advances in deep learning and computer vision as well as the increasing dependence on line-of-sight links in millimeter wave (mmWave) and terahertz systems. By leveraging vision, this new research direction enables an interesting set of new capabilities such as vision-aided mmWave beam and blockage prediction, proactive hand-off, and resource allocation among others. These capabilities have the potential of reliably supporting highly-mobile applications such as vehicular/drone communications and wireless virtual/augmented reality in mmWave and terahertz systems. Investigating these interesting applications, however, requires the development of special dataset and machine learning tasks. Based on the Vision-Wireless (ViWi) dataset generation framework \cite{ViWiDataset}, this paper develops an advanced and realistic scenario/dataset that features multiple base stations, mobile users, and rich dynamics. Enabled by this dataset, the paper defines the vision-wireless mmWave beam tracking task (ViWi-BT) and proposes a baseline solution that can provide an initial benchmark for the future ViWi-BT algorithms. 
\end{abstract}

\section{Introduction} \label{sec:Intro}

% Motivating vision-aided wireless
Wireless communication systems are moving towards millimeter wave (mmWave) frequency bands in 5G and  above 100GHz in 6G and beyond \cite{Andrews2014,6GAndBeyond}.  While the large available bandwdith at these high freqyency bands enables ultra high data rate communication, the features and propagation characteristics of the mmWav and terahertz signals introduce a set of challenges for these communications systems. First, mmWave (and terahertz) systems need to deploy large antenna arrays and use narrow beams to gurantee sufficient received signal power. Finding the best narrow beamforming vectors, however, requires large training overhead. Second, these high frequency signals experience severe penetration loss which renders blockages as a critical challenge for the wireless network coverage. These challenges (among others) make it difficult for mmWave and terahertz systems to relaibly support highly-mobile applications such as wireless virtual/augmented reality and vehicular communications. Interestingly, however, as the wireless systems move to higher frequency bands, the communication links between the transmitters and receivers become shorter with (visual) line-of-sight. This motivates leveraging visual data captured for example by RGB/depth cameras or Lidar to help overcome key challenges such as mmWave beam and blockage prediction---realizing what we call \textit{vision-aided wireless communications} \cite{CamBeamPred,ViWiDataset}.

%Motivation the development of new datasets and tasks 
Vision-aided wireless communication can find interesting applications in vehicular/ drone communications, virtual/ augmented reality, internet-of-things, gaming, and smart-cities among others. This exciting new direction of research is further motivated by (i) the advances in deep learning and computer vision, and (ii) the increasing interest in leveraging machine learning tools to address key wireless communication challenges such as mmWave and massive MIMO beam/channel prediction \cite{Li2019,CoordBeamForm,DetChPred,zhang2019deep}, blockage prediction and proactive hand-off \cite{Alkhateeb2018a,Alrabeiah2019a,mismar2018partially}, and intelligent resource allocations and user schduling \cite{Mismar19,8743390,8680025}.  Enabling the machine learning research in the vision-aided wireless communication direction requires the development of special datasets with co-existing visual and wireless data. With this motivation, we developed the Vision-Wireless dataset generation framework (ViWi) \cite{ViWiDataset}, which is a fremwork for generating co-existing and prametric datasets for wireless data (such as communication and radar channels) and visual data (such LiDAR as RGB/depth images). In the first ViWi release \cite{ViWiDataset}, we presented 4 ViWi scenarios/datasets that can enable a set of machine learning tasks/applications such as mmWave beam and blockage prediction as shown in \cite{CamBeamPred}. The scenarios adopted in the initial ViWi release, however, were relatively simple compared to practical outdoor and indoor environments.

To advanvce the research in a wider range of vision-wireless applications (machine learning tasks), we develop in this paper an advanced, detailed, and realistic ViWi scenario/dataset that incorporates multiple base stations and mobile objects/users (cars, buses, and pedestrians) featuring rich dynamics. Enabled by this dataset, we define the new machine learning task  \textit{ViWi vision-aided mmWave beam tracking} (ViWi-BT) which targets predicting future mmWave beams of mobile users using a sequence of previous beams and visual images. Leveraging vision has the potential of allowing beyond 5G transceivers to develop efficient 3D scene understanding \cite{song2015sun,zia2015towards,handa2016understanding}. In the context of the ViWi mmWave beam tracking task, this could enable the prediction not only of the future line-of-sight (LOS) beams of mobile users but also the non-LOS beams. Motivated by these interesting capabilities and their potential gains in enabling mmWave communiation in highly-mobile applications, and for the goal of encouraging more research in this direction, the ViWi-BT machine learning challenge \cite{ViWi-BT} is organized based on the dataset and task defined in this paper. Finally, for the sake of defining an initial benchmark for the  beam predicition accuracy, we present a baseline solution for the ViWi-BT task based on gated recurrent neural networks. This baseline solution uses only previously selected beams to predict future beams (i.e., without using the visual data). Leveraging the visual data in addition to the beam sequences is therefore expected to yield better prediction accuracies compared to those reported in the paper.

\begin{figure*}[h]
	\centering
	\subfigure[Top view of the visual instance.]{\includegraphics[width=0.45\textwidth]{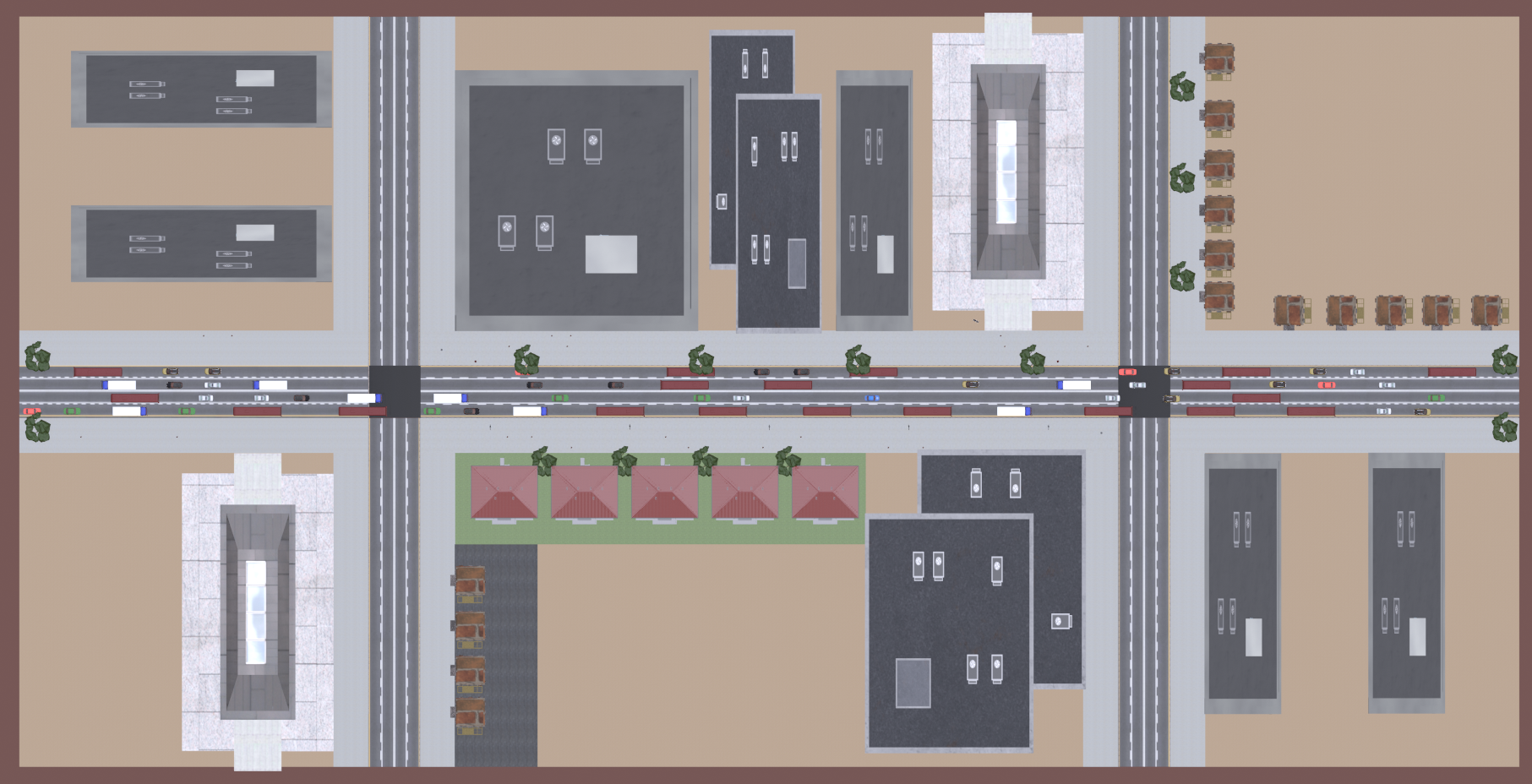}}
	\subfigure[Top view of the wireless instance.]{\includegraphics[width=0.45\textwidth]{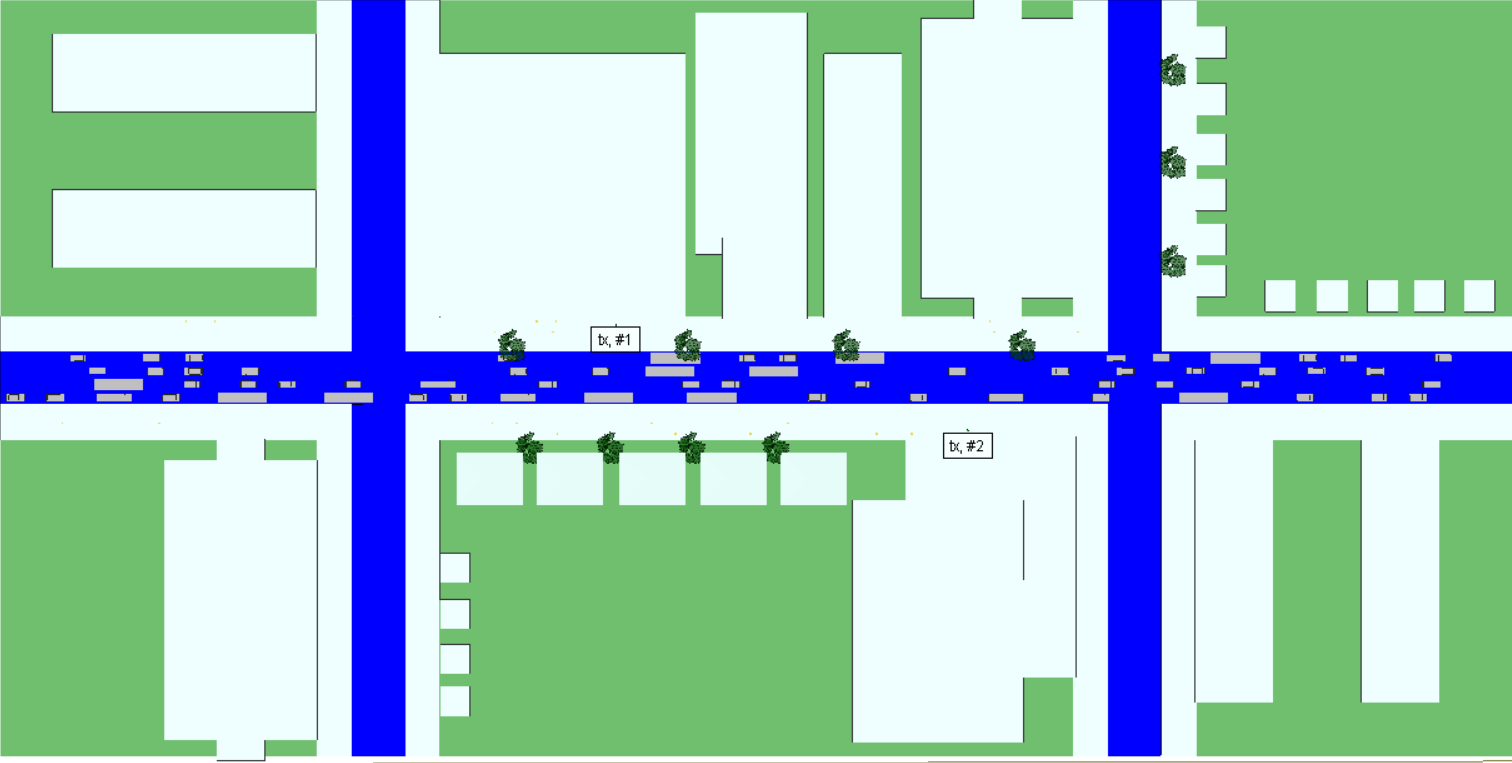}}
	\subfigure[Close-up from the visual instance.]{\includegraphics[width=0.45\textwidth]{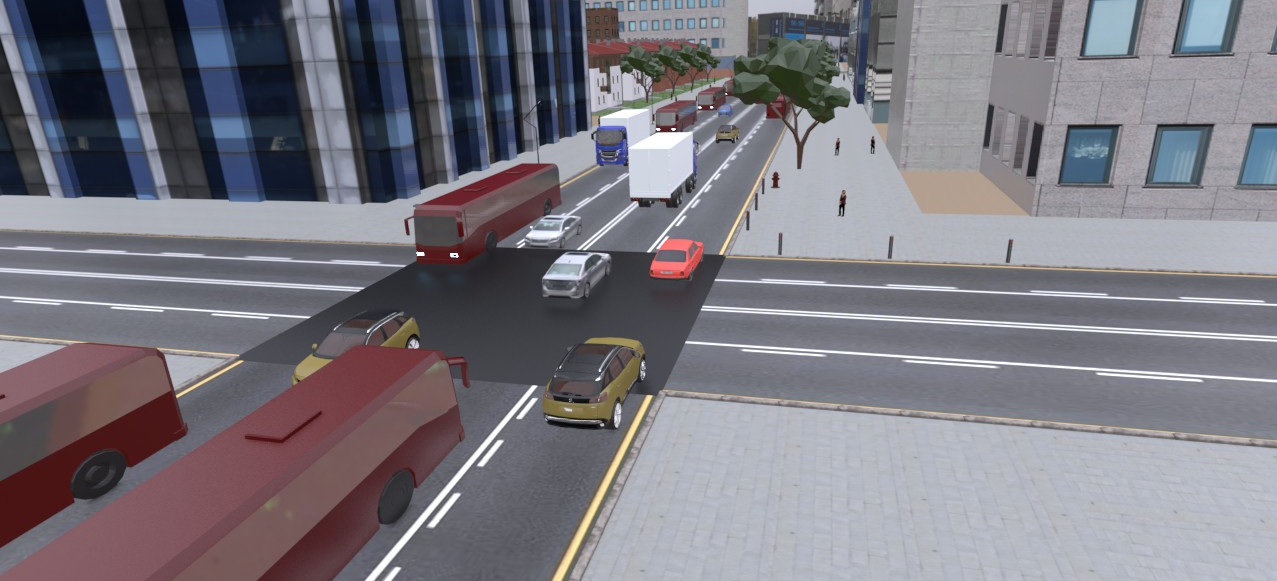}}
	\subfigure[Close-up from the wireless instance.]{\includegraphics[width=0.45\textwidth]{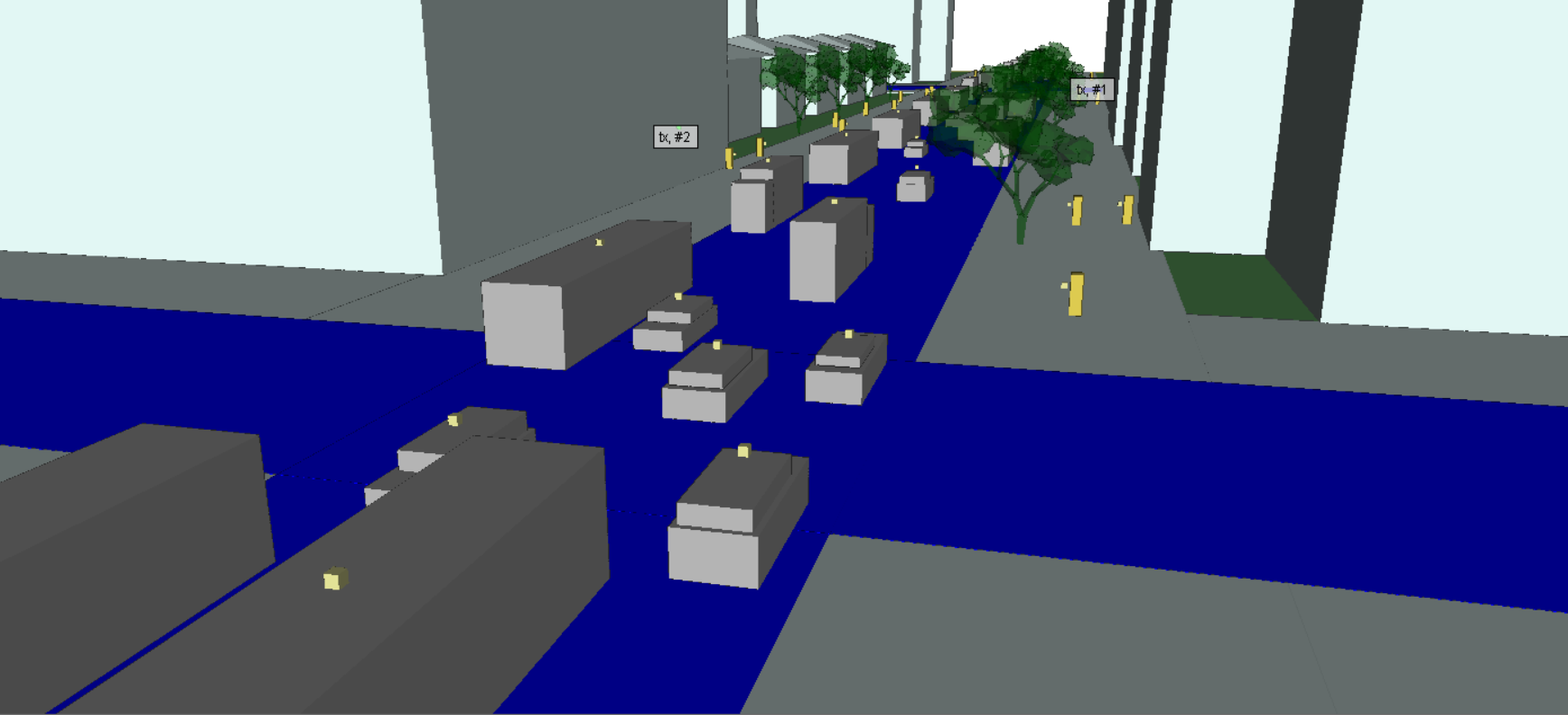}}
	\caption{Images of the visual and wireless simulation environments. This shows the difference between the two environments. The top row depicts top view of the whole environment showing the layout of the street. The bottom row shows two close-ups from the same scene that illustrate the two-block models used to in the wireless instance to replace the complex car models in the visual instance.}
	\label{top_view}
\end{figure*}

\section{Multi-User ViWi Scenario and Dataset} \label{sec:scenario}
Using the ViWi framework \cite{ViWiDataset}, a new scenario is designed specifically for the ViWi-BT task. It features multiple moving objects, constituting possible receivers, blockages, and reflectors. The following three subsections introduce a description of the scenario, seed dataset, and the development dataset.

\subsection{Scenario Description}
Different to previous ViWi scenarios \cite{ViWiDataset}, the new scenario depicts a dynamic environment showing busy traffic with pedestrians in a detailed downtown scape. The street, car, and human models used in the visual instance of the scenario, as shown in Fig. \ref{top_view}-a, are designed and rendered using the popular open source computer graphics and animation software Blender\textsuperscript{\textregistered}. This scene contains many commonly found objects in real world cities creating a complex scenario that contains static objects such as trees, bushes, sidewalks, benches, and buildings; as well as dynamic objects such as cars, buses, trucks, and people.
 
The 3D models of the scenario buildings, sidewalks, and streets are imported into the ray tracing simulator Wireless InSite\texttrademark \cite{Remcom}. The 3D models of the dynamic objects have very fine details with a high polygon count that creates unnecessarily high computation burden. To simplify the simulation, the dynamic objects are replaced with two-block approximations with no major impact on the accuracy of the simulation results. Fig. \ref{top_view}-b visualizes how the models were simplified for the simulation. The electromagnetic properties of each object are selected to correspond to the simulation frequency of 28 GHz. Table \ref{tbl:objects} shows detailed information about the objects used in the simulations.

\begin{table*}[t]
	\centering
	\caption{A list of objects composing the visual and wireless environments of the scenario}
	\begin{tabular}{c c c c}
		\hline
		\hline
		Object			&	Dimensions(Width, Length, Height in meters)		&	Material				&	Note	\\
		\hline
		\hline
		Road			&	215.05 x 429.88 x 0.50							& 	Concrete				&	-		\\
		Sidewalk		&	215.05 x 429.88 x 0.50							& 	Asphalt					&	-		\\
		Tree			&	7.58 x 8.87 x 9.95								& 	Dense Deciduous Forest	&	-		\\
		Street lamp		&	0.40 x 1.55 x 6.16								&	-						& 	Removed	\\
		Fire hydrant	&	0.62 x 0.62 x 1.15								&	-						&	Removed	\\
		Trash can		&	0.60 x 0.60 x 1.09								&	-						&	Removed	\\
		Street bollard	&	0.17 x 0.17 x 1.09								&	-						& 	Removed	\\
		Building 1		&	22.02 x 74.80 x 103.33							& 	Concrete				&	Replaced with same dimension block model	\\
		Building 2		&	43.43 x 91.11 x 130.33							&	Concrete				&	Replaced with same dimension block model	\\
		Building 3		&	63.35 x 87.11 x 74.34							&	Concrete				&	Replaced with same dimension block model	\\
		Building 4		&	32.24 x 87.11 x 74.34							&	Concrete				&	Replaced with same dimension block model	\\
		Building 5		&	69.75 x 74.80 x 54.21							&	Concrete				&	Replaced with same dimension block model	\\
		Apartment 1		&	19.21 x 19.38 x 11.48							&	Concrete				&	Replaced with same dimension block model	\\
		Apartment 2		&	10.88 x 10.18 x 20.58							&	Concrete				&	Replaced with same dimension block model	\\
		Truck			&	10 x 3 x 4										&	Metal					&	Replaced with two-block model	\\
		Bus				&	14 x 3 x 3										&	Metal					&	Replaced with same dimension box	\\
		Car 1			&	4.98 x 2 x 1.44									& 	Metal					&	Replaced with two-block model	\\
		Car 2			&	4.39 x 2.23 x 1.22								&	Metal					&	Replaced with two-block model	\\
		Car 3			&	4.29 x 1.82 x 1.31								&	Metal					&	Replaced with two-block model	\\
		Car 4			&	4.77 x 2.25 x 1.74								&	Metal					&	Replaced with two-block model	\\
		Car 5			&	4.56 x 2.09 x 1.68								&	Metal					&	Replaced with two-block model	\\
		Car 6			&	4.22 x 1.77 x 1.44								&	Metal					&	Replaced with two-block model	\\
		Car 7			&	4.93 x 2.14 x 1.38								&	Metal					&	Replaced with two-block model	\\
		Car 8			&	5.03 x 1.87 x 1.33								&	Metal					&	Replaced with two-block model	\\
		Human 1			&	0.43 x 0.47 x 1.54								&	Human Skin				&	Replaced with same dimension box	\\
		Human 2			&	0.68 x 0.43 x 1.86								&	Human Skin				&	Replaced with same dimension box	\\
		\hline
		\hline
	\end{tabular}
	\label{tbl:objects}
\end{table*}

The new scenario comprises multiple scenes, each of which is one pair of visual and electromagnetic instances. To simulate real traffic, all objects in the scene are set with their own speed and are initialized randomly at a certain trajetory. A full description of the different elements of the environment are described below:
%A scene contains 100 moving objects each with separate trajectories. These trajectories were generated with a custom traffic simulator which approximates the behavior of real cars and pedestrians. The scene consists of 4 lanes of traffic with a sidewalk on either side of the road, leaving four car lanes and four pedestrian lanes. Each moving object is initialized in a randomly selected lane with a randomly generated speed selected uniformly from the minimum and maximum speeds designated for that object type. After initialization, objects travel down their respective lane until they reach the end of the lane. They change speed when they encounter a vehicle traveling at a slower speed along the same lane. This guarantees that objects with different speeds do not collide. When an object reaches the end of the lane in the simulation, it is reinitialized in another randomly selected lane with a new speed.

%The scenario contains two base-stations positioned on opposite sides of the street at either end of the scenario. Each base station is equipped with a transmitter antenna and three cameras pointed in non-overlapping positions to incorporate a 180 degree field of view. Each dynamic object possesses a receiver antenna. The antenna for the human models are located on their side at waist height to represent a phone in their hand or pocket. The receivers for the cars, trucks, and buses are located on the top of the respective object similar to a vehicle antenna.

\begin{itemize}
	\item\textbf{Layout}
	This is an outdoor scenario of a metropolitan area, here shown from a top view perspective in Fig.\ref{top_view}-(a) and -(b). There is one main (horizontal) street and two secondary streets which are lined by a sidewalk on either side. The secondary streets form two four-way intersections that separate the scene into six sections where buildings are placed. The length of the main street is 429.88 meters and the secondary streets are 215.05m. The width of both street types is 15m. Along each street is a sidewalk that is 10m wide.
	\item\textbf{Buildings}
	For this model, office and apartment buildings are placed along the sidewalk to depict a city street. There are five different building types and two apartment building types to add realism to the scene. The dimensions of these structures vary and are available in Table I for reference. Referring to the top view figures (a) and (b): Apartment building 1 and 2 are placed in the top right and bottom middle sections of the scene. Building 1 is in the bottom right and top left sections. Building 2 is placed in the bottom left and top middle section. Buildings 4,5,6 are in the top middle section and building and building 3 is in the bottom middle. 
	
	\item \textbf{Dynamic objects:} At any instance during the scenario, 100 dynamic objects with a diverse number of models exist in the scene. These include 20 human models, 35 - 50 car models, 5 - 10 truck models, and 25 - 35 bus models. Each object possesses a unique trajectory which approximates the behavior of a real car or pedestrian. Each moving object is initialized in a randomly selected lane with a randomly generated speed selected uniformly from the minimum and maximum speeds designated for that object type. Cars, trucks, and buses have initial speeds ranging from 10 to 20 m/s, while human models have initial speeds between 1 and 2 m/s. After initialization, objects travel down their respective lane until they reach the end of the lane. They change speed when they encounter an object traveling at a slower speed along the same lane. The change in speed is governed by one simple rule, maintaining a minimum distance with other objects. For vehicles, this distance is set to 5 meters while for humans, it is set to 1 meter. This guarantees that objects with different speeds do not collide. When an object reaches the end of a lane in the simulation, it is reinitialized in another randomly-selected lane with a new speed. For the wireless instance of each scene, every dynamic object is fitted with a half wave dipole receiver antenna oriented along the z axis. The antenna for the human models are located on the side of the object to simulate a phone in their hand or pocket. The receivers for the cars, trucks, and buses are located on the top of the respective object at a length of approximately 30 cm.
	
	\item \textbf{Base stations:} The scenario contains two base-stations positioned on opposite sides of the street at either end of the main street, spaced 60 meters apart.  Each base station is equipped with a half wave dipole antenna array oriented along the z axis with 128 antennas along the x axis. Each base station also includes three cameras pointed in different directions such that they cover the whole street. 
\end{itemize}

\subsection{The Seed Dataset}
\label{sec:dataset}
%(This should describe, briefly, the use of ViWi framework to generate wireless channels before we discuss the final sturcture of the development dataset.)
Following the generation pipeline of ViWi \cite{ViWiDataset}, an initial Vision-Wireless dataset is generated. The pipeline goes through three main stages to generate pairs of RGB images and 28 GHz wireless channels. They make up the seed to the development dataset. The generation process is itemized as follows:
\begin{itemize}
  \item \textbf{Scenario Definition:} since the new scenario has multiple moving objects traveling at different speeds, the scenario definition consists of multiple scenes that share the same stationary and dynamic visual and electromagnetic elements, but they differ in where each dynamic object is located.
  \item \textbf{Raw-Data Generation:} similar to the previously-released ViWi scenario, the scenario is fitted with raw-data generators. They are the same across all scenes, and they are used to generate different raw-data sets, one per scene.
  \item  \textbf{Parameterized Processing:} this stage converts the raw visual and wireless data into a set of pairs of RGB images and wireless channels \cite{ViWi}. This set will henceforth be referred to as the \textit{seed dataset}.
 \end{itemize}

\subsection{The Development Dataset: Image-Beam Sequence Dataset}\label{dev_data}
The seed dataset is further processed to generate the final development dataset, referred to as Image-Beam Sequence dataset. Since the focus in this paper is on the task of ViWi-BT, the development dataset must comprise sequences of pairs of beamforming vectors and RGB images. The beamforming vectors are obtained from a fixed beam-steering codebook, and, therefore, they could be identified by their indices. Using the wireless channels of each user, the optimal beams from the codebook are identified over a sequence of consecutive scenes, in which the user is changing position, and they are paired with the corresponding images depicting the user in each scene. An important point to emphasize here is that the dataset does not include localization information for the users depicted in each image, e.g., there are no bounding boxes identifying which user is served by which beamforming vector. More technical information on the Beam-Image dataset could be found in the README.txt file enclosed with the dataset package \cite{ViWi}.

%\clearpage
\section{Vision-Aided Beam Tracking} \label{VABT}
A problem of significant interest in mmWave communications is the selection and tracking of beamforming vectors. The dynamics of wireless environments place heavy computational burden on mmWave transceivers, either at the base station or user equipment side, to select and track the best beamforming vectors \cite{Sub6PredMmWave}\cite{CoordBeamForm}. This calls for the development of proactive and relatively light-weight methods that determine the best beamforming vectors ahead of time. Those methods should, by nature, be adaptive and display a form of intelligence, which suggests that machine learning should be their driving force.

For the purpose of encouraging the development of machine-learning algorithms for beam tracking, the ViWi-BT identifies the problem of mmWave beam tracking using wireless and visual data as its main task. The following two subsections will introduce the considered system and channel models and state the definition of the beam-tracking problem.

\subsection{System and Channel Model}
Next, we summarize the  mmWave system and channel models adopted in this work. 

 \label{sec:sys_model}
\textbf{System Model}: the communication system considered for the ViWi-BT task assumes two different transceiver setups. The first is for the base stations, and it is a fully-analog mmWave transceiver equipped with an $M$-element uniform linear array. The other is for the user equipment, and it is a single-antenna transceiver. The system is assumed to operate in a downlink mode and under Orthogonal Frequency-Division Multiplexing (OFDM). The received signal of the $u$th user equipment at the $k$th subcarrier could be give as:
\begin{equation}\label{sig_mod}
  y_{u,k} = \mathbf h_{u,k}^T \mathbf f_u x_{u,k} + n_{u,k}, 
\end{equation}
 where $x_{u,k}$ is the transmitted signal of the $u$th user at the $k$th sub-carrier, $\mathbf f_u \in \mathbb C^M$ is the $M$-dimensional beamforming vector selected from codebook $\mathcal F$ for the $u$th user, $\mathbf h_{u,k} \in \mathbb C^M$ is the  channel vector at the $k$th sub-carrier between the $u$th user and the base station, and $n_{u,k} \sim \mathcal N(0,\sigma^2)$ is an AWGN sample.
 
 \textbf{Channel Model:} a geometric channel model is assumed for the tracking problem, in which the channel vector of the $i$th user is given by:
 \begin{equation}
\mathbf{h}_{u,k} = \sum_{d=0}^{D-1} \sum_{\ell=1}^L \alpha_\ell e^{- \j \frac{2 \pi k}{K} d} p\left(dT_\mathrm{S} - \tau_\ell\right) \ba\left(\theta_\ell, \phi_\ell\right),
\end{equation} 
where $L$ is number of channel paths, $\alpha_\ell, \tau_\ell, \theta_\ell, \phi_\ell$ are the path gains (including the path-loss), the delay, the azimuth angle of arrival, and the elevation angle of arrival, respectively, of the $\ell$th channel path. $T_\mathrm{S}$ represents the sampling time while $D$ denotes the cyclic prefix length (assuming that the maximum delay is less than $D T_\mathrm{S}$). 

\subsection{Problem Definition}
\label{sec:prob_def}
Proactively predicting future downlink beamforming vectors based on previously-observed beamforming vectors and images of the environment is the core task of ViWi-BT. First, we consider user $u$ and define the  \textit{sequence} of its observations  at time instance $t \in \mathbb{Z}$ as 
\begin{equation}
\mathcal{S}_u[t] = \{(\mathbf f_u[t-r+1], X[t-r+1]), \dots, (\mathbf f_u[t],X[t])\},
\end{equation}
where $\mathbf f_u[m]$ is the beamforming vector in codebook $\mathcal F$ used to serve user $u$ at time instance $m, m \in \mathbb{Z}$. $X[m]\in \mathbb R^{W\times H\times C}$ is an RGB image of the environment in which the $u$th user appears at the $m$th time instance \footnote{$W$, $H$, and $C$ are, respectively, the width, hight, and color dimensions of the image.}, and $t,r\in \mathbb Z^{+}$ are respectively the current time instance and the extent of the observation interval, the objective is to identify the best $N$-future beamforming vectors  that maximize the achievable rate of the future $N$ instances. Formally, the objective is to identify the set of optimal beams $\mathcal{P}^\star_u[t]=\{\mathbf f^\star_u[t+1],\dots,\mathbf f^\star_u[t+N]\}$, with $\bff^\star_u[m]$ defined as
\begin{equation}\label{opt_beam}
  \bff_u^\star[m] = \underset{\mathbf f \in \mathcal{F}}{\text{argmax}} \sum_{k=1}^K  \log_2\left(1+ \mathsf{SNR} \left\|\mathbf{h}_{u,k}[m]^T \mathbf{f}_{u}\right\| _2^2 \right),
\end{equation}
where $\mathbf{h}_{u,k}[m]$ is the channel vector betweem user $u$ and the base station at instance $m$ and subcarrier $k$, and $\mathsf{SNR}$ denotes the SNR. 
The task of identifying the set $\mathcal{P}_u^\star[t]$ requires a prediction algorithm. Such algorithms in machine learning could be described as a prediction function $f_{\Theta}(\mathcal{S})$ parameterized by a set of parameters $\Theta$. This function is not expected to be customized for a single user, but it should be capable of predicting the set $\mathcal{P}^\star_u$ of any user in the wireless environment, whether it is a LOS or NLOS user. In other words, the prediction function should maintain high overall prediction accuracy. Mathematically, this is expressed as follows: if $\hat {\mathcal{P}}_u[t]$ is the prediction of $f_{\Theta}(\mathcal{S}[t])$ for user $u$, then $f_{\Theta}(\mathcal{S}[t])$ should have high probability of correct prediction given the observed sensory data $\mathbb P\left(\hat {\mathcal{P}}_u[t] = \mathcal{P}_u^{\star}[t]| \mathcal{S}_u[t] \right)$, and over all users $U$ in the wireless environment \footnote{Generally, this number is approximated by a statistical sample size. This sample forms the training dataset, and it is expected to be generated using the random process that governs the dynamics of the environment \cite{DLBook}}.

\section{Beam Tracking Using Beam Sequences}
\label{sol_1}
A valid and important question that could be posed at this stage is whether there is a need for visual sensory data to tackle the beam tracking problem. This section attempts to answer this question and motivate the need for visual data from an empirical perspective; it introduces a Recurrent Neural Network (RNN) that is designed to perform future beam prediction based on previously-observed beam sequences. Below are descriptions of the proposed model and its training procedure followed by a discussion on its performance.

\begin{figure}[hbt]
  \includegraphics[width=\linewidth]{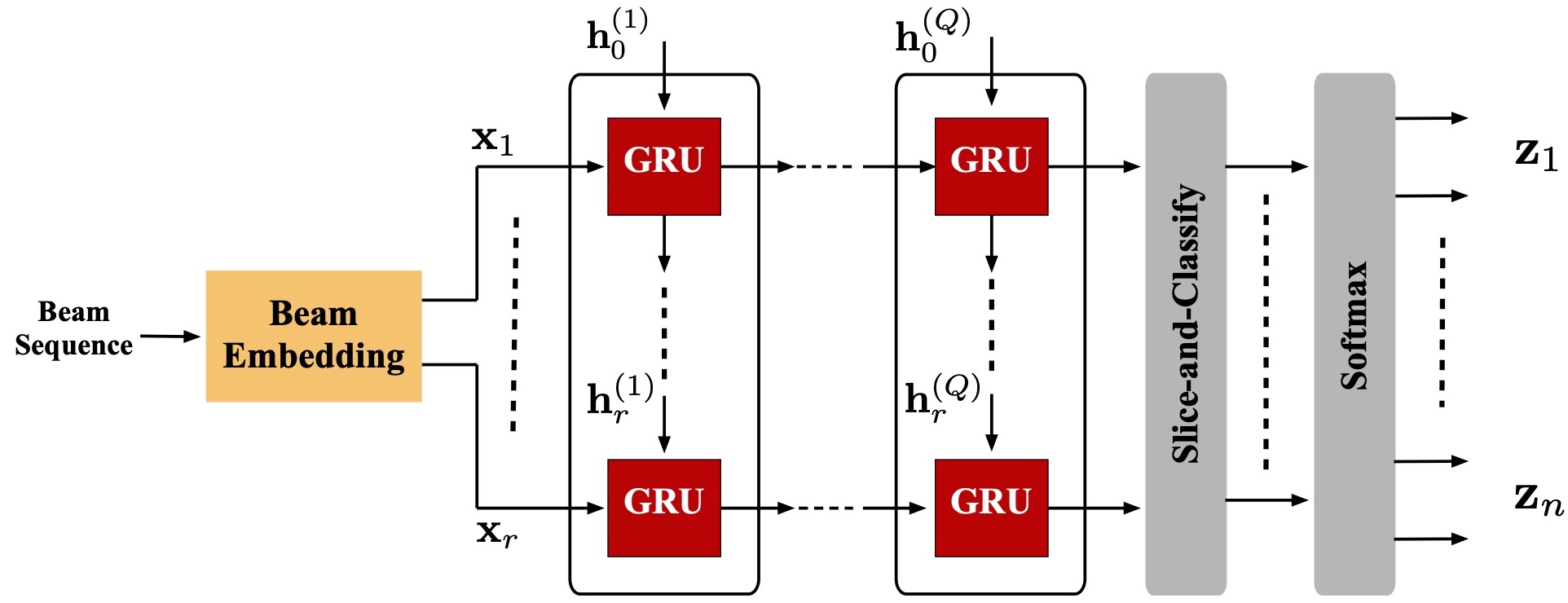}
  \caption{This block diagram depicts the architecture of the $Q$-layer recurrent neural network with Gated Recurrent Units (GRU) and a slice-and-classify layer. It takes in an embedded sequence of $r$ beams and produces a sequence of $N$ probability vectors $\mathbf z_n, n=1, ..., N$.}
  \label{arch}
\end{figure}

\subsection{Network Architecture}
Fig.\ref{arch} shows the general architecture of the proposed RNN. It is a vanilla architecture, comprising $Q$ recurrent layers and a slice-and-classify layer. Each of the recurrent layers implements $r$ Gated Recurrent Units (GRUs). They have been mainly chosen because of their good performance combating the short-term memory issue \cite{DLBook} as well as the good performance they demonstrated in tackling similar proactive prediction tasks, \cite{BlockagePred} to name an example. They are also interesting baseline models as they are less complex than the state-of-the-art Long Short-Term Memory (LSTM) networks.

An important design parameter of the above-described model is its depth (number of layers $Q$); depth has been shown to have a great impact on increasing the expressibility of neural networks while maintaining a relatively manageable number of parameters compared to wide shallow models, see \cite{DLBook}\cite{VGG}\cite{ResNet}\cite{ExpPowOfDepth}. As a result, it has been chosen to be the main variable to study the effectiveness of using beam sequences to do beam tracking. 

Three different-depth vanilla models are explored, namely 2-, 4-, and 6-layer networks. All of them share the same input, hidden state, and output dimensions as well as the input pre-processing stage, see Table \ref{model_trn}. The inputs are sequences of length $r$ of best beam indices read from the dataset described in Section \ref{dev_data}. Since these indices are just integers, each one needs to be transformed into a real-valued vector $\mathbf x_{i_r}\in \mathbb R^D$ $\forall {i_r}\in \{1,\dots,r\}$, with $D$ denoting the embedding dimension. This is usually referred to as input embedding. Such transformation is the only pre-processing procedure applied to the input data. At the output end of the network, the slice-and-classify layer takes all the outputs of the last $Q$th layer, and cut out the last $N$ of them to feed to the softmax layer. The softmax, then, produces $N$ vectors of probabilities $\mathbf z_n\in \mathbb R^{|\mathcal F|} \ \forall n\in\{1,\dots,N\}$. The index of the element with the highest probability in every vector $\mathbf z_n$ is the index of the predicted future beam $b[t+n] \in \mathbb Z^+ \  \forall n \in \{1,\dots,N\}$. The scripts implementing and training these models are available at \cite{MyGitHub}.

\begin{table}
	\centering
	\caption{Design Parameters and Training Hyper-parameters}
	\begin{tabular}{c|cc}
	\hline\hline \Tstrut
	\multirow{7}{*}{Design}
	& Length of input sequence ($r$) & 8 \\ 
	& Prediction sequence length ($N$) & 1, 3, and 5 \\
	& Embedding dimension ($D$) & 50 \\
	& hidden state dimension & 20 \\
	& Codebook size ($|\mathcal F|$) & 128 \\
	& Depth & 2, 4, and 6 \\
	& Dropout percentage & 0.2 \\
	\hline \Tstrut
	\multirow{4}{*}{Training}
	& Solver & ADAM\cite{ADAM} \\
	& Learning rate & $1\times 10^{-3}$ \\
	& Batch size ($B$) & 1000 \\
	& Number of epochs & 100 \\
%	&  & \\

	\hline\hline
	\end{tabular}
	\label{model_trn}
\end{table}

\subsection{Network Training}
All three models are trained on the training set of the ViWi-BT dataset. As the target is to study the ability of a machine learning model to perform beam tracking using sequences of previously-observed beams, only beam indices from the training set are used, no images. Each model is trained three times, each of which is for a longer prediction sequence, i.e., different $N$ lengths. Table \ref{model_trn} shows the lengths and the training hyper-parameters. The training is conducted with a \text{cross entropy} loss \cite{DLBook} given by
\begin{align}
  L = \frac{1}{N}\sum_{n= 1}^{N} \left\{ \sum_{j = 1}^{|\mathcal F|} q_{nj}\log{z_{nj}} \right\},
\end{align}
averaged over the batch size, where $\mathbf z_n = [z_{n1},\dots,z_{n|\mathcal F|}]^T$ and $q_n= [q_{n1},\dots,q_{n|\mathcal F|}]^T \in \{0,1\}^{|\mathcal F|}$ are, respectively, the output probabilities and the target one-hot vector\footnote{A binary vector with one element having the value of 1} of the $n$th predicted beam.

\subsection{Performance Evalustion}

In this section, we evaluate the performance of the proposed beam tracking solution. 

\textbf{Evaluation metrics:} following the training of each model, it is tested on the validation set of ViWi-BT. Two metrics are defined to evaluate the performance of a model. First one is the top-1 accuracy  \cite{Sub6PredMmWave}\cite{CamBeamPred}. It is defined as:
\begin{equation}
  Acc_{\text{top-1}} = \frac{1}{A}\sum_{i = 1}^{A} \mathbbm{1}\{ \hat {\mathcal{G}}_i = \mathcal{G}_i^{\star} \},
\end{equation}
where $\hat{\mathcal{G}}_i$ is the set of predicted beam indices of the $i$th data sample with length $n$, $\mathcal{G}_i^{\star}$ is the groundtruth best beams of the same data sample, $A$ is the total number of validation data samples, and $\mathbbm{1}\{.\}$ is the indicator function with value of 1 only when the condition is met. The groundtruth beam indices are obtained from the solution of Eq.\ref{opt_beam}. The second metric is the \textit{exponential-decay score}, which is given by:
\begin{equation}\label{score}
\mathrm{score} = \frac{1}{A} \sum_{i=1}^A \exp\left(-\frac{||\mathbf {\hat g}_i-\mathbf g_i^{\star}||_1}{n \ \sigma}\right),
\end{equation}
where $\mathbf {\hat g}_i$ and $\mathbf g^{\star}_i$ are, respectively, vectors of the predicted and optimal beam indices (vector form of $\hat G_i$ and $G^{\star}_i$) of the $i$th data sample, and $\sigma$ is a penalization parameter.

The two metrics serve different purposes. Top-1 accuracy is very popular in classification problems \cite{ImageNet12}\cite{VGG}\cite{AlexNet}. However, it is a hit-or-miss type of metric, which may not be fair for beam tracking. When a future beam is mis-identified, that does not necessarily mean all non-optimal beams are equivalent; some non-optimal beams have better performance than others. Hence, the second metric is designed to address the \textit{gradual} impact of mis-identification. It treats the predictions and targets (labels) as vectors, so it can measure the first norm distance between the two. This distance is scaled by a penalization term and plugged into an exponential kernel. This results in a smoother and more flexible evaluation metric than top-1 accuracy.

\begin{figure}[hbt]
  \includegraphics[width=\linewidth]{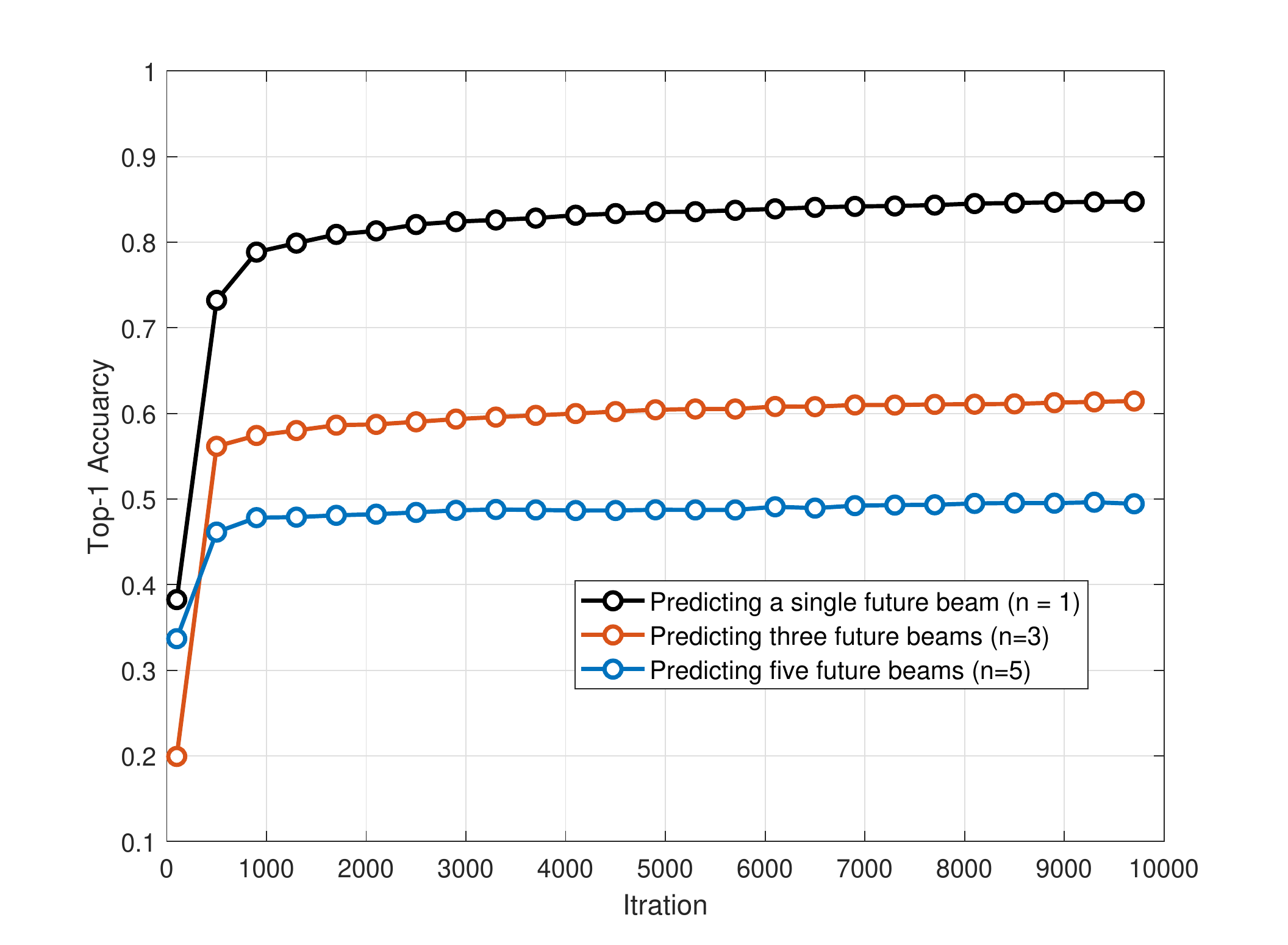}
  \caption{This shows the improvement of Top-1 accuracy of the RNN in Section.\ref{sol_1} as training progresses. The accuracy is reported as a ratio of the correctly predicted sequences to the total number of validation sequences.}
  \label{perf_top}
\end{figure}

\textbf{Prediction Performance:} Fig.\ref{perf_top} plots the validation top-1 accuracy versus number of training iterations for the 2-layer model; it is the model with best performance among the three. The figure shows some interesting trends. When the task is to predict the next future beam, the RNN performs well and achieves an 85\% top-1 accuracy. However, once the number of future beams starts increasing, the model starts to struggle. Its top-1 accuracy crumbles down from 85\% to around 60\% and 50\%, respectively, as the length grows to 3- and 5-future beams. Same trend could be seen in Table.\ref{expo_score} for the exponential score with a penalization parameter $\sigma=0.5$. Another interesting observation from the figure is the rate at which the performance improves. The model learns the beam patterns very quick in all cases, but it also saturates quickly when the number of future beams is greater than 1. 

\begin{table}
	\centering
	\caption{Exponential Decay Scores}
	\begin{tabular}{c|c|c}
	\hline\hline
	
	Single future beam & Three future beams & Five future beams \\
	\hline
	86\% & 68\% & 60\% \\
	
	\hline\hline
	\end{tabular}
	\label{expo_score}
\end{table}

The aforementioned observations, although not definitively, point towards an interesting conclusion; only observing beam sequences do not reveal enough information about future beam sequences. This conclusion could be solidified by the results of the deeper models, which are expected to be more expressive \cite{ExpPowOfDepth}. For the case of predicting 5 future beams, the most challenging case, the deep models do not improve upon the performance of the 2-layer model. In fact, they register slightly worse performance than that of the shallow model. The 4-layer model saturates at around 48.61\% top-1 accuracy while the 6-layer hits only 48.15\% top-1 accuracy. 

These results suggest that extra sensory data like images of the surrounding environment could constitute a great supplementary source of information to deal with the beam tracking problem; images are rife with information about the environment, and with the right machine leaning algorithm, they could be utilized to identify possible future blockages and possible reflecting surfaces as well as to understand users' motion patterns. All are things that are hard to do, if not impossible, with only observed beam sequences.

%-------------------------------------------------%
\section{Conclusion}\label{concl}
Taping into the potential of mmWave and massive MIMO communications requires developing innovative communication paradigms such as those characterized by artificial intelligence. Vision-aided wireless communications  is one great example of such paradigms at which visual data (captured for example by RGB/depth cameras) are leveraged to address the key wireless communication challenges. The objective of this paper is to facilitate the research and development of vision-wireless solutions by  introducing an advanved ViWi dataset/scenario as well as defining the new ViWi vision-aided mmWave beam tracking task (ViWi-BT). The developed ViWi scenario models a dynamic outdoor mmWave communication environment, comprising thousands of visual and wireless instances (scenes). This scenario undergoes the ViWi data-generation process \cite{ViWiDataset} to produce a collection of images and wireless channels called the seed dataset. The collection is further processed to generate a development dataset that consists of hundreds of thousands of data samples, each of which is a sequence of image and beam pairs. This dataset enables the development of machine learning based solutions for several important tasks such as mmWave beam tracking leveraging both wireless and visual data. To draw some insights into the potential of the developed dataset and to provide an initial benchmark, we presented a baseline solution that only uses previous beam sequences to predict future beams. The beam prediction accuracy is expected to improve significantly by developing innovative solutions leveraging both wireless and visual data, which is an interesting future research direction. 

%-------------------------------------------------%
\balance
\bibliographystyle{IEEEtran}

\begin{thebibliography}{10}
\bibitem{ViWi-BT}
	\BIBentryALTinterwordspacing
	[Online]. The link will be posted here soon! 
	\BIBentrySTDinterwordspacing
\end{thebibliography}

%, AlkhateebRef.bib
\end{document}